\documentclass[11pt]{article}

\usepackage[hyphens]{url}
\usepackage{pdflscape}
\usepackage[margin=1in]{geometry}
\usepackage[toc]{appendix}
\usepackage{graphicx}
\RequirePackage[OT1]{fontenc}
\usepackage{natbib}
\usepackage{adjustbox}
\usepackage{caption}
\usepackage{subcaption}
\usepackage{amsmath,amssymb,amsthm,bm,latexsym}
\usepackage{graphicx,psfrag,epsf}
\usepackage{epigraph,xcolor}
\usepackage{enumerate}

\usepackage{booktabs}
\usepackage{caption}
\usepackage{multirow}
\usepackage{hyperref}
\usepackage{float}

\usepackage{marginnote}
\usepackage{pseudocode}
\usepackage{algpseudocode}
\usepackage[linesnumbered,ruled]{algorithm2e}
\hypersetup{colorlinks=true,citecolor=blue,urlcolor=purple,
	pdfpagemode=FullScreen,linktocpage=true}
\usepackage{cleveref}

\usepackage{tikz,lipsum,lmodern}
\usepackage[most]{tcolorbox}

\allowdisplaybreaks

\numberwithin{equation}{section}

\theoremstyle{definition}
\newtheorem{exmp}{Example}[section]

\theoremstyle{remark}
\newtheorem{rem}{Remark}[section]

\usepackage{tikz,xcolor,hyperref}

\definecolor{lime}{HTML}{A6CE39}
\DeclareRobustCommand{\orcidicon}{%
	\begin{tikzpicture}
	\draw[lime, fill=lime] (0,0) 
	circle [radius=0.16] 
	node[white] {{\fontfamily{qag}\selectfont \tiny ID}};
	\draw[white, fill=white] (-0.0625,0.095) 
	circle [radius=0.007];
	\end{tikzpicture}
	\hspace{-2mm}
}

\foreach \x in {A, ..., Z}{%
	\expandafter\xdef\csname orcid\x\endcsname{\noexpand\href{https://orcid.org/\csname orcidauthor\x\endcsname}{\noexpand\orcidicon}}
}

\begin{document}

\title{Model Selection for Ordinary Differential Equations: a Statistical Testing Approach}

\author{Itai Dattner\orcidA{}
\\Department of Statistics, University of Haifa
\\idattner@stat.haifa.ac.il
\and
Shota Gugushvili\orcidB{}
\\Biometris, Wageningen University \& Research
\\shota.gugushvili@wur.nl
\and
Oleksandr Laskorunskyi\orcidA{}
\\Department of Statistics, University of Haifa
\\olaskoru@campus.haifa.ac.il
}

\maketitle

\begin{abstract}
Ordinary differential equations (ODEs) are foundational in modeling intricate dynamics across a gamut of scientific disciplines. Yet, a possibility to represent a single phenomenon through multiple ODE models, driven by different understandings of nuances in internal mechanisms or abstraction levels, presents a model selection challenge. This study introduces a testing-based approach for ODE model selection amidst statistical noise. Rooted in the model misspecification framework, we adapt foundational insights from classical statistical paradigms (Vuong and Hotelling) to the ODE context, allowing for the comparison and ranking of diverse causal explanations without the constraints of nested models. Our simulation studies validate the theoretical robustness of our proposed test, revealing its consistent size and power. Real-world data examples further underscore the algorithm's applicability in practice. To foster accessibility and encourage real-world applications, we provide a user-friendly Python implementation of our model selection algorithm, bridging theoretical advancements with hands-on tools for the scientific community.

\end{abstract}

\section{Background and Motivation}\label{sec:backmot}
\subsection{Mechanistic Modelling with Ordinary Differential Equations}\label{sec:genback}
Differential equations have proven to be a powerful modeling tool in science and engineering. They are widely used for modelling purposes, e.g., in mathematical biology, see \cite{edelstein2005mathematical} and \citep{murray02}; in the theory of chemical reaction networks, see \cite{feinberg1979lectures}; in biochemistry, see \cite{voit2000computational}; and in compartmental models in epidemiology, see \cite{anderson1992infectious}.

On an abstract level, differential equations
comprise a class of mechanistic models. Mechanistic models are typically developed based both on the empirical knowledge and on the fundamental laws of the nature (first principles), and harness some information on causal mechanisms governing a system of interest. Upon their calibration, mechanistic models can be leveraged in applications where experiments are either impossible or costly to perform, ideally yielding new and valuable insights into a phenomenon under study, cf. \cite{baker_2018}, and leading to better prediction and control of dynamic processes, cf. \cite{strogatz2018nonlinear}.

\subsection{Aims and Contribution}\label{sec:aims}

In many situations one wants to compare several ODE models for a given phenomenon. This multiplicity of models arises, e.g., when some internal mechanisms governing the process are known only approximately. Other times there is contradictory scientific knowledge on underlying causal relationships, resulting in differing ODE model formulations. Finally, a possibility to choose the abstraction or resolution level at which to represent mathematically a given phenomenon may also lead to competing ODE models. In fact, often the detail level in ODE modelling is dictated by computational and feasibility considerations. See \cite{dattner2017modelling} and \cite{vanvoorn2023} for some examples of the above considerations.

In this paper, we focus on the common and practically important case of complex dynamic processes observed with statistical noise. Extensive lists of references on statistical modelling and inference for dynamical systems that cover a wide selection of areas can be found in \cite{ramsay2006functional} and \cite{ramsay2017dynamic}. For a recent review of the role played by differential equations in data analysis, with a focus on parameter estimation for ODE models, see \cite{dattner2021differential}.

Under this statistical setup, we study model selection issues. In particular, we devise an applicable methodology that sheds additional light on the modelling questions at hand and that has a potential to translate into practical recommendations or actions.

\subsection{Approaches to Model Selection for ODEs}
\label{sec:approaches}

This section reviews the model selection problem for ODEs, while drawing on general literature on model selection, such as \cite{ripley04} and \cite{wit12}. Simultaneously, it provides some motivation on our take on model selection for ODEs.

To avoid possible misunderstanding, we recall that a good explanatory model is not always the best predictive model (\cite{shmueli10}, \cite{perretti13}, \cite{hartig13}, and \cite{perretti13bis}). The primary goals of ODE modelling are to explain the phenomenon under study and make predictions from the model, see \cite{murray02}. For prediction, alternative approaches to mechanistic modelling, e.g. machine learning methods or discrete time models may also be considered (\cite{michailidis13}, \cite{lindsey01}, \cite{ellner98}, \cite{kendall99}, and \cite{thakur91}). Our work targets the explanatory behavior of ODE models, and as such assumes that their simplicity, robustness, and basis in natural laws will result in reasonable predictive behavior. Rigorous study of the latter, that may involve cross-validation techniques, is a topic on its own.

Our approach is based on the following premises, that will be illustrated and expanded upon below.

\begin{tcolorbox}[colback=blue!5!white,colframe=blue!75!black]
\begin{itemize}
    \item ODE models are a stylised representation of reality. They are not the `truth'.
    \item Scientific knowledge may dictate more than one ODE model for a given phenomenon. None of these models is `true'.
    \item Only a handful of ODE models needs to be compared at any time.
    \item Penalty-based statistical model selection approaches for ODEs require resolution of a number of conceptual challenges.
    \item Information-theoretic criteria do not provide rigorous assessment of statistical significance of the model selection results.
\end{itemize}
\end{tcolorbox}

The first statement above hardly warrants a discussion: the fact that ODE models are derived based on simplifying assumptions and as such cannot be thought to be `true' is widely acknowledged in the modelling literature; cf. \cite{murray02} and \cite{lindsey01}.

\begin{exmp}
    The simplest growth model for bacterial population under controlled conditions (constant temperature, sufficient supply of the nutrient, and others) in the lab environment is given by the Malthus law:
    \begin{align*}
x^{\prime}(t) & = \psi x (t), \\
x(0) & = \xi.
    \end{align*}
    Here $x(t)= \xi \exp(\psi t)$ is the bacterial density at time $t$, $\xi$ is the initial value, and $\psi$ is the growth constant; see, e.g., \cite{edelstein2005mathematical}, Section 4.1. Derivation of this equation is based on several simplifying assumptions, e.g.\ that $x$ is sufficiently large so that addition of several individuals to the population is of negligible consequence, growth of individuals is not correlated, and death can be neglected (\cite{edelstein2005mathematical}, p.~117). None of these assumptions can be thought as absolutely true in all circumstances. Yet the model has been proven to be adequate in practice under specific conditions; see, e.g., \cite{edelstein2005mathematical}, p.~120.
\end{exmp}

Typically, ODE model selection arises when there are competing explanations for a phenomenon under study.

\begin{exmp}
\label{ex:dattner17}
    \cite{dattner2017modelling} study the following model for interaction of two bacterial populations, the predatory bacterium \emph{Bdellovibrio bacteriovorus} and its prey, \emph{Burkholderia stabilis st.2.}, in a lab experiment:
    \begin{align*}
        x_1^{\prime}(t) & = \psi_1 \psi_2 x_3(t) - \psi_3 x_1(t),\\
        x_2^{\prime}(t) & = -\psi_4 (x_2(t) - \psi_5) x_1(t),\\
        x_3^{\prime}(t) & = \psi_4 (x_2(t) - \psi_5) - \psi_2 x_3(t).
    \end{align*}
    Here $x_1$, $x_2$ and $x_3$ are concentrations of the predator, prey and the predator-prey complex (bdelloplast), respectively. The complex  is not observed and is introduced into the ODE system to account for the fact that the time it takes the predator to handle its prey is of the same order as the time it takes for the consumed prey items to be converted into new predators (\cite{dattner2017modelling}, page 2). An alternative here could have been some form of the classical Lotka-Volterra model, which only involves the $x_1$ and $x_2$ components. The role of the refuge parameter $\psi_5$ is interesting, in that it models the fact that due spatial inhomogeneities, at any given time not all prey are available for predation to the predator. It goes without saying that reduction of the spatial effects to a single parameter $\psi_5$ is a serious simplifying assumption. See. e.g., \cite{edelstein2005mathematical}, pp.~87--89 for an additional discussion.
\end{exmp}

The best model needs to be chosen based on scientific knowledge, available data, and model complexity, see \cite{ripley04}. Statistics provides a suitable and principled formalism for informed decision-making for model selection. In turn, statistical model selection encompasses model discrimination and model testing; see, e.g. \cite{fisher_mcaleer79}, \cite{mcaleer_bera83}, \cite{dastoor81}, and \cite{dastoor90} for additional details. According to \cite{dastoor90}, model testing arises when it is desired to test the `truth' of a model of interest. On the other hand, model discrimination applies when two or more models are ranked and compared according to some criterion. The latter can be `deterministic'\footnote{Strictly speaking, the term `deterministic' is a misnomer. We interpret it loosely as a model discrimination approach that does not involve a significance test.} and based on quantities like information criteria, Mallows's $C_p$ and adjusted $R^2$, or `probabilistic' and involve a significance test.

At present, there is limited literature focussing specifically on model selection for ODEs. Important references include, among others, \cite{miao2009differential}, \cite{zhang15}, and \cite{wu2019parameter}. For Bayesian methodologies, see, e.g., \cite{girolami08}, \cite{girolami11}, \cite{oates16} and \cite{hug16}. A systematic study of scientific papers from 1990--2023 revealed that out of 91 articles discussing `model selection' and `differential equations', approximately 60\% mentioned information criteria, 22\% mentioned Bayesian-like criteria, and 25\% mentioned cross-validation criteria. Besides these, at least 15 novel methods were introduced to deal with the selection of the `best' dynamical system describing a given natural phenomenon (see Supplementary Material).

In general, for model discrimination one can use information criteria, e.g. the Akaike Information Criterion (AIC) and the Bayesian Information Criterion (BIC), see \cite{akaike73}, \cite{akaike1974new} and \cite{schwarz1978estimating}. However, the expert opinions are divided as to which criterion, if any, is the most suitable in practice. Next, AIC and BIC rely on the trade-off between the goodness-of-fit term and the penalty term penalising model complexity. The latter is difficult to interpret for competing ODE models, that may be nonnested, have a differing number of state variables, differ in severity of their nonlinearity, and involve different external forcing functions or covariates.

\begin{exmp}
    Revisit Example \ref{ex:dattner17}. The classical Lotka-Volterra model and the model proposed in \cite{dattner2017modelling} have differing numbers of state variables (two and three, respectively). How the two models should be nested within a single all-encompassing ODE model is unclear.
\end{exmp}

\begin{exmp}
    In Section \ref{subsect:real_exam2} ahead, several Lotka-Volterra type systems are considered to model interaction of two populations: predators and their preys. Each one attempts to improve upon the basic Lotka-Volterra system by addressing one of its unrealistic consequences based on the knowledge of the phenomenon under study. Relative weights of these consequences cannot be objectively assessed by simply counting the corresponding parameters. In the linear regression setting, prior to model selection the covariates or features are standardized. This allows a fair assessment of their relative contributions to the response. The same tool is not available in the ODE setting. Cf.\ the discussion in \cite{vissing2017}, pp.~6 and 29.
\end{exmp}

We note in passing that the experts disagree as far as applicability of information criteria to model selection for nonnested statistical models is concerned; cf.\ \cite{ripley04} and \cite{burnham02}.

Importantly, model discrimination approaches such as AIC and BIC are criticised for not providing probabilistic quantification of the significance of their results; see \cite{vuong89} and \cite{amemiya80}. In fact, it is often the case that AIC and BIC assign similar scores to several models, and while one model is still chosen as the best, the fact whether in probabilistic terms it is significantly better than its competitors remains elusive. Various rules of thumb used in practice to assess statistical significance of relative differences between competing models based on information criteria (see, e.g., \cite{burnham02}) lack formal theoretical justification and moreover are not universally applicable.

\subsection{Overview of our approach}

A general alternative to `deterministic' model discrimination is testing. It is the approach we employ for the model selection problem for ODEs. The aim is to leverage peculiarities of the ODE modelling in a targeted way and do not treat the question as a routine model selection problem. Importantly, we adopt the model misspecification framework, where the researcher's principal goal is finding the best possible explanation of the data-generating mechanism given a parametric model; see, e.g., \cite{cox61}, \cite{cox62}, \cite{white1981consequences} and \cite{vuong89}.

Our testing approach builds upon \cite{vuong89}, that in turn has its roots in \cite{hotelling40}. Importantly, it does not require artificial nesting of competing models. As already mentioned in Section \ref{sec:approaches}, in the context of ODE model selection, nesting is not always possible, and at any rate it leads to consideration of additional and irrelevant models, thereby resulting in entirely avoidable computational difficulties and the increased computational cost; see \cite{zhang15}, \cite{ramsay2007parameter} and \cite{voit04} for a discussion of computational issues associated with parameter estimation in ODE models. With testing, the case of more than two models can be handled through pairwise comparison, taking suitable  care of multiple testing issues; cf. \cite{schennach17}. The nature of the ODE modelling is such that one typically needs to compare only a handful of competing models, so that eventual multiple testing corrections do not result in overly conservative tests. Since with Vuong's test nesting is not a requirement, one can consider, compare and rank genuinely different causal explanations of an empirical phenomenon. Note that here we do not discuss the network reconstruction problems using ODEs, as studied, e.g., in \cite{henderson2014network} and \cite{chen2017network}. Though the distinction is not entirely clear-cut, the latter are in their character closer to exploratory analysis, whereas the problems we have in mind lean towards confirmatory analysis; cf. \cite{snedecor1989}, p. 64.

Vuong's approach starts with the null hypothesis that the two models are equally close to the true data-generating mechanism, which is not required to be contained among either of the competing models. The alternatives to the null hypothesis are that either the first or the second model is closer to the true data-generating distribution. Testing is based on the likelihood ratio statistic, and the test is directional. Critical values for Vuong's test are obtained from the asymptotic distribution of the likelihood ratio statistic, which varies depending on whether the models are nested or non-nested. Determining nestedness for two ODE models can be difficult, and is in fact oftentimes impossible. The latter difficulty in Vuong's approach is bypassed via the use of a pre-test. 

However, such a two-step approach to testing can lead to a considerable size distortion (\cite{shi15}, \cite{schennach17}). Schennach and Wilhelm's modification of Vuong's approach, referred to as the S-W test, addresses the issue of test size distortion and works regardless whether the models are nested or not; see \cite{schennach17}.

Testing approaches in \cite{vuong89} and \cite{schennach17} were not developed with ODE models in mind, but aimed at classical statistical and econometric models. In the ensuing sections we demonstrate how their work can be adapted to the ODE framework. The extension is nontrivial, but we show that all the necessary details can be worked out. 

\section{Real Data Examples}
Prior to delving into technical details, in order to give a taste of how our testing approach works, in this section we present results of its application on several real data examples.

\subsection{Agricultural Trial Data}
\label{subsect:real_exam1}
\begin{figure}
\centering
\includegraphics[width=8cm]
{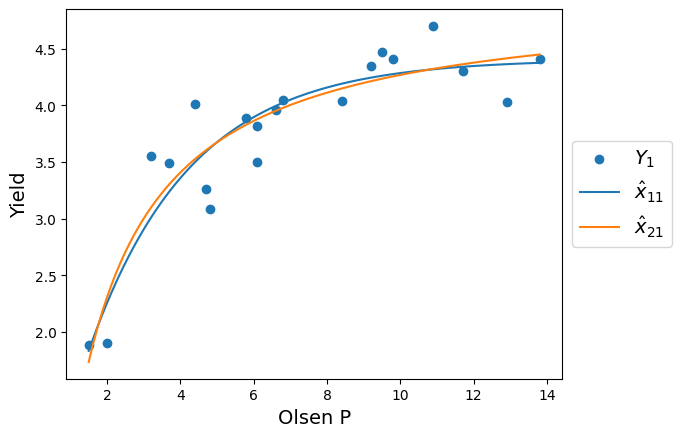}
\caption{Relationship between the yield and the soil phosphorus content, and fits of two ODE models from Section \ref{subsect:real_exam1}. The first index of the variable $x$ indicates the model, that is the first one and the second one. The second index indicates the state (single one in this case).}
\label{fig:exam_1}
\end{figure}

\cite{welham2014statistical}, Example 17.1A, provide data of a field trial at Rothamsted Research that studied the relationship between crop yields and applications of the soil fertilizer. The data are yields of spring barley from 20 fields in 1986, and the available soil phosphorus content, measured as Olsen P\footnote{The dataset is available online at \url{http://www.stats4biol.info/wordpress/examples/data/} under the name \texttt{PHOSPHORUS.DAT}.}. We plot the data in Figure \ref{fig:exam_1}. 

\cite{welham2014statistical}, Example 17.1C, propose several models for functional relationship between the (mean) yield and phosphorus, of which we focus on the standard exponential model with an asymptote (equation (17.7) in \cite{welham2014statistical}) and the inverse linear model (equation (17.8) there). Upon rewriting these 3-parameter nonlinear functions in the ODE form and reparametrizing, we obtain two ODE models:
\begin{align*}
\marginnote{{\color{blue} Exponential}}
x^{\prime}(s) & = \psi_1 (\psi_2 - x(s)),\\
x(0) & = \xi,
\end{align*}
and
\begin{align*}
\marginnote{{\color{blue} Inverse linear}}
x^{\prime}(s) & = -\psi_1 (-\psi_2 + x(s))^2, \\
x(0) & = \xi,
\end{align*}
The state variable $x$ is yield and is a function of the phosphorus content $s$. Furthermore, $\psi_1, \psi_2$ are the model parameters, and $\xi$ is the initial value.

To fit the models, we used the nonlinear least squares method. This is essentially the maximum likelihood estimation approach under the assumption of Gaussian measurement errors. As seen in Figure \ref{fig:exam_1}, visually both models fit the data well. Nevertheless, the two models can be rigorously compared via the testing framework that we develop in Sections \ref{subsect:models} and \ref{subsect:method} below. In our software implementation, little beyond reading the data in and specifying the ODE models via straightforward syntax is required from the user. The outcome of the test statistic is $-0.359$. Under the commonly used significance level $\alpha=0.05$, the critical value to reject the null hypothesis that both models are equally distant (in the Kullback-Leibler divergence sense) from the true data-generating mechanism in favor of model B is $-1.96$, and in favor of model A is $1.96$. As our test statistic is between these two  values, we retain the null hypothesis.

\cite{welham2014statistical} compare the models based on $R^2$, AIC and BIC, and conclude that ``there is little statistical difference in the fit of the two non-linear models, so either might reasonably be selected'' (page 475). Our results formally corroborate their conclusions.

\subsection{Paramecium Aurelia vs Saccharomyces Exiguus Data}
\label{subsect:real_exam2}

This example deals with results of a lab experiment on interaction of two species: \emph{Paramecium aurelia} (predator) and \emph{Saccharomyces exiguus} (prey). For details of the experiment, we refer to \cite{gause35experimental}. We obtained data by means of WebPlotDigitizer (see \cite{rohatgi2022}) by reading the point coordinates of Figure 3 in  \cite{gause35experimental}. The graphical quality of the original figure is modest, and hence we rounded off the measurement readings to one digit after zero.

\begin{figure}
     \centering
     \begin{subfigure}[b]{0.48\textwidth}
         \centering
         \includegraphics[width=\textwidth]{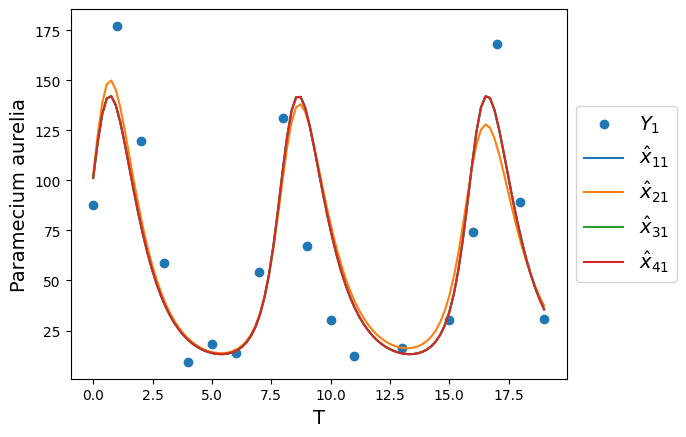}
        \caption{Predator (\emph{Paramecium aurelia})}
         \label{fig:exam2_pred}
     \end{subfigure}
     \hfill
     \begin{subfigure}[b]{0.48\textwidth}
         \centering         \includegraphics[width=\textwidth] {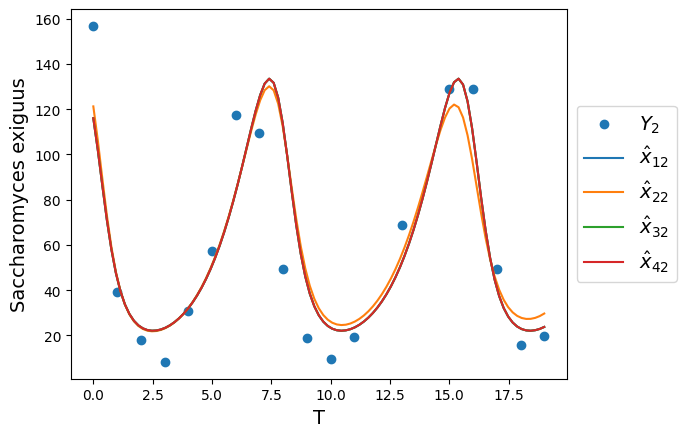}
         \caption{Prey (\emph{Saccharomyces exiguus})}
         \label{fig:exam2_prey}
     \end{subfigure}
        \caption{Dynamics of the predator and prey populations, with four fitted ODE models from Section \ref{subsect:real_exam2}. Note that as some fits are nearly identical, not all four lines are visible.  The first index of the variable $x$ indicates the model. The second index indicates the state.}
        \label{fig:exam2}
\end{figure}

As seen from Figure \ref{fig:exam2}, both populations exhibit characteristic oscillations with an approximately constant period and lend themselves to modelling via a Lotka-Volterra type ODE system. We used the following ODE model formulations:

\begin{itemize}
\item Model 1:
\begin{align*}
\marginnote{{\color{blue} Lotka-Volterra}}
x_1^{\prime}(t) & = \psi_2\psi_3x_1(t)x_2(t)-\psi_4x_1(t),\\
x_2^{\prime}(t) & =\psi_1x_2(t)-\psi_2x_1(t)x_2(t),\\
\textbf{x(0)} & = [\xi_1, \xi_2];
\end{align*}
\item Model 2:
\begin{align*}
\marginnote{{\color{blue} Logistic prey}}
x_1^{\prime}(t) &= \psi_2\psi_3x_1(t)x_2(t)-\psi_4x_1(t),\\
x_2^{\prime}(t) &=\psi_1x_2(t)\left(1-\frac{x_2(t)}{\psi_5}\right)-\psi_2x_1(t)x_2(t),\\
\textbf{x(0)} & = [\xi_1, \xi_2];
\end{align*}
\item Model 3:
\begin{align*}
\marginnote{{\color{blue} Type 2 functional response}}
x_1^{\prime}(t) & = \frac{\psi_2\psi_3x_1(t)x_2(t)}{1+\psi_2\psi_5x_1(t)} - \psi_4x_1(t), \\
x_2^{\prime}(t) & =\psi_1x_2(t) -\frac{\psi_2x_1(t)x_2(t)}{1+\psi_2\psi_5x_1(t)}, \\
\textbf{x(0)} & = [\xi_1, \xi_2];
\end{align*}
\item Model 4:
\begin{align*}  
\marginnote{{\color{blue} Density-dependent predator death}}
x_1^{\prime}(t) & = \psi_2\psi_3x_1(t)x_2(t) - \psi_4x_1(t)-\psi_5x_1^2(t),\\
x_2^{\prime}(t) & =\psi_1x_2(t) - \psi_2x_1(t)x_2(t),\\
\textbf{x(0)} & = [\xi_1, \xi_2].
\end{align*}
\end{itemize}

Here Model 1 is the basic Lotka-Volterra model, whereas each of the successive ones attempts to address one of its shortcomings. For instance, under the basic Lotka-Volterra model, the prey population can increase exponentially. This is reasonable for the prey at low density. However,   in real populations as the density becomes higher, the per-head rate of increase declines. Model 2 attempts to account for this by introducing a logistic limitation on the prey growth. In a similar fashion, Model 4 lets the predator vitality rate be a function of the predator density. This appears reasonable, in that predators lacking territories might start infighting or suffer higher death rate. For additional information on each model, we refer to \cite{murdoch2003}; cf. \cite{edelstein2005mathematical}, pp.~214--217.

In the present context, results of model selection can be interpreted in a twofold fashion. Firstly, we may ask whether there is enough information in the observed time series to discern the refinements of the basic Lotka-Volterra model. Secondly, we may ask which of these extensions is statistically the most significant.

Figure \ref{fig:exam2_pred} shows the dynamics of \emph{Paramecium aurelia} ($Y_{1}$) and the four fitted models (the first number in the subscript of $\hat{x}_{11},\ldots,\hat{x}_{41}$ refers to the model and the second number to the state). In the same manner, Figure \ref{fig:exam2_prey} gives the dynamics of \emph{Saccharomyces exiguus} and the respective fits. Visually the model fits appear to be similar enough. Results of our testing procedure are reported in Table \ref{table:exmp2_tests}. The table implies that no model is shown to be closer to the true data-generating process than others. No multiple testing correction has been applied when presenting the results, as none of the pairwise comparisons turned out to be significant at the conventional $\alpha=0.05$ level.

\begin{table}
\captionsetup{font=small}
\caption{Test outcomes for Models 1--4 in Section \ref{subsect:real_exam2}.}
\begin{center}
\small
    \centering
    \begin{tabular}{c c c c} 
    \toprule 
     {\bf Model A} & {\bf Model B} & {\bf S-W statistic} & {\bf In favor}\\
    \midrule
1 & 2 & 0.987 & $-$\\
1 & 3 & 0.791 & $-$\\
1 & 4 & 0.680 & $-$\\
2 & 3 & -0.706 & $-$\\
2 & 4 & -0.706 & $-$\\
3 & 4 & 0.727 & $-$\\
    \bottomrule
    \end{tabular}
\end{center}
\label{table:exmp2_tests}
\end{table}

Upon a closer look at the four ODE systems, we find that Model 2 equals Model 1 when $\psi_5=\infty$, and Models 3 and 4 equal Model 1 when $\psi_5=0$. It is instructive to examine the estimated parameters shown in Table \ref{table:exmp2_estimates}. We see that parameter estimates for Models 1, 3 and 4 are nearly identical, and in Models 3 and 4 $\psi_5\approx0$. The same goes for Model 2, except that $\psi_5$ is now large. The latter is not surprising, given that $\psi_5$ plays a different role in Models 3 and 4, on one hand, and Model 2 on the other.

\begin{table}
\captionsetup{font=small}
\caption{Parameter estimates for Models 1--4 in Section \ref{subsect:real_exam2}.}
\begin{center}
\small
    \centering
    \begin{tabular}{c c c c c } 
    \toprule
     {\bf Parameter} & {\bf Model 1} & {\bf Model 2} & {\bf Model 3} & {\bf Model 4}\\
    \midrule
$\xi_1$ & 101.2& 102.8& 101.2 &101.2\\
$\xi_2$ & 116.0& 121.2& 116.0&116.0\\
$\psi_1$ &0.660&0.685 &0.660 &0.660\\
$\psi_2$ &0.012&0.012 &0.012 &0.012\\
$\psi_3$ &1.450&1.531 &1.490 &1.490\\
$\psi_4$ &1.122&1.126 &1.122 &1.122\\
$\psi_5$ & $-$   &1566.4 &5.93E-15 &3.10E-18\\
    \bottomrule
    \end{tabular}
\end{center}
\label{table:exmp2_estimates}
\end{table}

\subsection{Paramecium Bursaria vs Schizosaccharomyces Pombe Data}
\label{subsect:real_exam3}
In the previous example, among four suggested predator-prey ODE systems, no model has shown statistically significant superiority over other models. Here we will show the case when the test works in favor of a model. We fitted the same models as in Section \ref{fig:exam2} to a different dataset, namely the dataset {\bf gause\_1934\_book\_f39.1} from the {\bf R}  package {\bf gauseR}, see \citet{muhlbauer2020gauser}. This deals with interaction of the predator \emph{Paramecium bursaria} and the prey \emph{Schizosaccharomyces pombe} (see \cite{gause35experimental} for details of the experiment).

\begin{figure}
     \centering
     \begin{subfigure}[b]{0.48\textwidth}
         \centering
         \includegraphics[width=\textwidth]{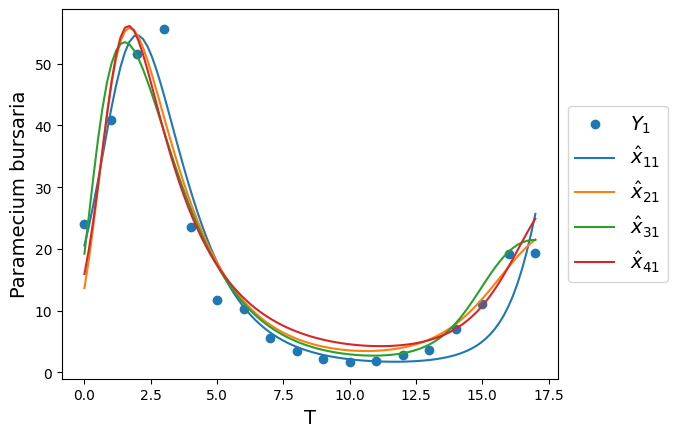}
        \caption{Predator (\emph{Paramecium bursaria})}
         \label{fig:exam3_pred}
     \end{subfigure}
     \hfill
     \begin{subfigure}[b]{0.48\textwidth}
         \centering         \includegraphics[width=\textwidth]{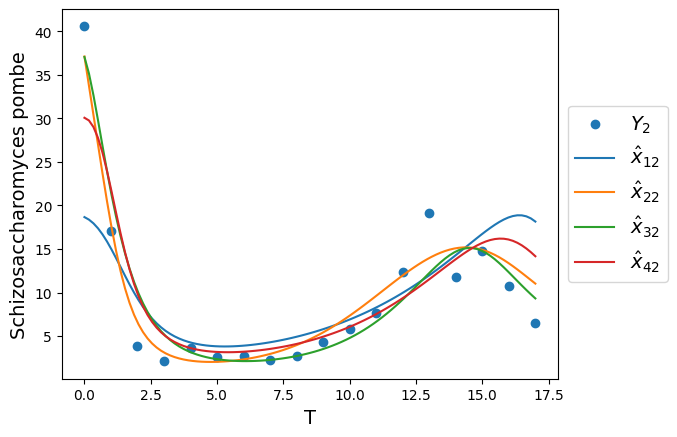}
         \caption{Prey (\emph{Schizosaccharomyces pombe})}
         \label{fig:exam3_prey}
     \end{subfigure}
        \caption{Dynamics of the predator and prey populations, with four fitted ODE models from Section \ref{subsect:real_exam2}.}
        \label{fig:exam3}
\end{figure}

Figure \ref{fig:exam3} gives the dynamics of the observed data and plots the fitted models. On purely visual grounds, it is hard to conclude which model is better. The test results are reported in \ref{table:exmp3_tests}. At the $95\%$ significance level, Model 2, which is `Logistic prey', is closer to the truth in comparison to Models 1 and 4 (see Table \ref{table:exmp3_tests} for details). In the pair of Models 2 and 3, the shift $0.908$ towards Model 2 is not enough to obtain a statistically significant result. Nevertheless, one may still prefer Model 2 over Model 3, because the latter does not achieve a statistically significant improvement over Models 1 and 4. In this case, the increased number of samples could have given additional information. At any rate, the dataset is somewhat unusual, in that the magnitude of periodic fluctuations in Figure \ref{fig:exam3} diminishes as the time progresses (compare the first and the second cycles). \citet{gause35experimental} notes this, but refrains from giving an explanation. This example illustrates well the challenges associated with ODE model selection.

\begin{table}
\captionsetup{font=small}
\caption{Test outcomes for Models 1--4 in Section \ref{subsect:real_exam3}.}
\begin{center}
\small
    \centering
    \begin{tabular}{c c c c} 
    \toprule 
     {\bf Model A} & {\bf Model B} & {\bf S-W statistic} & {\bf In favor}\\
    \midrule
1 & 2 & -4.433 &2\\
1 & 3 & -1.827 &-\\
1 & 4 & -1.374 &-\\
2 & 3 &  0.908 &-\\
2 & 4 &  5.802 &2\\
3 & 4 &  1.680 &-\\
    \bottomrule
    \end{tabular}
\end{center}
\label{table:exmp3_tests}
\end{table}

\section{Problem Formulation}
\label{subsect:models}

We start by introducing a number of concepts and notions to formally define our statistical framework.

\subsection{Statistical Modeling}\label{sec:obsetup}
As we make a distinction between the true data generating process and ODE-based approximations to it, we first need to discuss the former.

We assume the data collected on the phenomenon of interest are pairs $(t_i, Y_{i})$,  $i=1,\ldots,n$. The $t_i$'s can typically be thought of as times, at which measurements $Y_i$'s are collected\footnote{As illustrated in Section \ref{subsect:real_exam1}, `time' is not the only possible interpretation of $t_i$'s.}. As a specific example, $Y_i$'s may represent some quantitative estimates of the number of individuals infected by a certain disease at times $t_i$'s. In this paper, for the sake of clarity of exposition, we suppose that $(t_i,Y_i)$ are independent and identically distributed random vectors and follow a common (unknown) probability distribution $P_0$ with density $p_0$. This setup can be generalised to a more abstract one, but we do not attempt this here. The assumption is flexible enough to cover numerous situations of practical interest; see Remark \ref{rem:iid} below. Our approach is thus a probabilistic approach to the description of empirical phenomena and is a point of view taken by researchers in statistics and related fields, such as econometrics and machine learning (see, e.g., \cite{wasserman04}, pp.~ix and 19, and \cite{white94}, pp.~5--6). The distribution $P_0$ (equivalently, its density $p_0$) gives a complete probabilistic description of the data-generating mechanism. A researcher's goal is inference on this unknown distribution.

Often the distribution of times $t_i$'s is of little relevance or can be assumed to be known (e.g., uniform), and in that case, the primary object of interest is the conditional distribution of $Y_i$ given $t_i$, say $P_0(\cdot | \cdot).$ We assume it has a density $p_0(\cdot | \cdot).$ The first step in inference on $P_0(\cdot | \cdot)$, or equivalently $p_0(\cdot | \cdot)$, is the formulation of a certain approximation to it, termed a statistical model. Next, in the second step, the model is optimised based on available observational or experimental data, and thereby the best approximation to $P_0(\cdot | \cdot)$ or $p_0(\cdot | \cdot)$ is obtained (\cite{white94}, Chapter 2).

The departure point for our statistical models is the following generic observational structure,
\begin{equation}\label{eq:obs}
Y_j(t_i)=x_{0j}(t_i)+\epsilon_{ij}, \quad i=1,\ldots,n \quad j=1,\ldots,d_0,
\end{equation}
where $Y_j(t_i)$ is a scalar random variable; $x_{0j}(t)$, $t\in[0,T]$ is an unknown deterministic function;  $t_1,\ldots,t_n$ are design points; and the unobserved random variables $\epsilon_{ij}$ are independent measurement errors having zero expectation and finite variance. Such modelling assumptions are standard in the literature dealing with statistical inference for ODEs systems; see, e.g., \cite{ramsay2007parameter}, \cite{hooker2009forcing}, \cite{gugu12},  and \cite{dattner15chris}, to name just a few references. We do not assume that \eqref{eq:obs} is the structure matching the `true' data generating distribution $P_0$. In particular, model misspecification can occur in the additive error assumption, distributional assumptions on the error terms, and the form of the mean function. Note that our derivations in further sections are under the assumption of Gaussian noise with zero mean and variance $\sigma^2_j$, yet the overall approach is general and can be adapted to alternative likelihood functions as well. 

Now for each $i$, we aggregate $Y_j(t_i)$'s into vectors $Y_i$'s and hence assume that the pairs $(t_i,Y_i)$ are independent and identically distributed.

\begin{rem}
    \label{rem:iid}
    Suppose that the observation times $t_i$ are independent and identically distributed, and furthermore that they are independent of measurement errors $\epsilon_{ij}$'s. Then the pairs $(t_i,Y_i)$ will be independent and identically distributed as well. In practice observation times $t_i$ are typically deterministic, for instance they could be daily. But if their empirical distribution stabilizes to a limiting distribution, $t_i$'s can be reasonably assumed to be independent and identically distributed. In particular, this is the case when $t_i$'s form a regular and dense grid on the time interval $[0,T]$. See, e.g. \cite{tsybakov2009introduction} for the use of similar ideas to establish asymptotic properties of nonparametric regression estimators in the fixed design setting, and compare to \cite{gasser1979}. 
\end{rem}

Let $\top$ stand for the transpose of a vector. In the sequel we use the notation
\[
x_0(t)=(x_{01}(t),\ldots,x_{0d_0}(t))^\top
\]
and denote the vector of derivatives of $x_0(t)$ w.r.t. $t$ by
\begin{equation}
\label{eq:f}
f_0(t)=(x^\prime_{01}(t),\ldots,x^\prime_{0d_0}(t))^\top, \quad t\in[0,T]. 
\end{equation}
The scientific question studied in this work is essentially a question of finding a parametric description for $f_0(\cdot)$ defined in Equation (\ref{eq:f}), one that expresses $f_0(t)$ in terms of $x(t)$ that describes the process mechanistically, in the sense that the current rate-of-change depends on the current state. Suppose that we have $N$ models for describing a dynamic process. We denote such models by $F_k$, $k=1,\ldots,N$ and assume with some innocuous abuse of notation that
\begin{equation}
\label{eq:ode}
\begin{cases}
x_k^{\prime}(t)= F_k(x_k(t),\psi_k), \quad t\in[0,T],\\
x_k(0)=\xi_k.
\end{cases}
\end{equation}
Here $x_k(0)=\xi_k$ is a column $d_k$-vector of initial conditions. The parameter vector is given by 
\begin{equation}
\label{eq:par}
\psi_k=(\psi_{k1},\ldots,\psi_{kp_k})^\top, 
\end{equation}
where $\psi_k$ is an element of a $p_k$-dimensional parameter space $\Psi_k$.

Let  
\[
x_k(t):=x(t;\xi_k, \psi_k)=(x_{k1}(t;\xi_k, \psi_k),\ldots,x_{kd_k}(t;\xi_k, \psi_k))^\top
\]
stand for the solution of the initial values problem defined by the system of ODEs and initial values given in Equation (\ref{eq:ode}) and a parameter $\psi_k$ defined in (\ref{eq:par}). Furthermore, define
\begin{equation}
\label{theta}
\theta=(\sigma^2, \xi, \psi)
\end{equation}
where $\sigma^2$ is the d-vector of variances of the noise of each state and we assume $\theta\in\Theta$ for some subset $\Theta$ of the Euclidean space.

Now the question is: given competing models $F_k$, which one should we prefer? The answer obviously is the one that gets us closest to $P_0$. Hence we need to discuss the distance between the true data generating distribution $P_0$ and the one implied by \eqref{eq:obs} for each competing model under consideration. We denote that latter distribution of the pair $(t_i,Y_i)$ by $P(\cdot,\cdot;{\theta})$, and assume that the marginal distribution of $t_i$'s is fixed, e.g.\ uniform on $[0,T]$.

As argued e.g.\ in \cite{white94}, pp.~9--10, and in \cite{akaike73}, Sections 2--3, specifically in the model discrimination context, a sensible and natural discrepancy measure between probability distributions $P$ and $Q$ is the Kullback-Leibler divergence (see \cite{kullback51})
\[
\operatorname{KL}(P,Q)=
\begin{cases}
\int \frac{\mathrm{d} P}{\mathrm{d} Q} \log\left( \frac{\mathrm{d} P}{\mathrm{d} Q} \right) \mathrm{d} P, & P \ll Q,\\
\infty, & \textrm{otherwise}.
\end{cases}
\]
When $P$ and $Q$ possess densities $p$ and $q$, as is our case, the Kullback-Leibler divergence can be equivalently written as
\[
\operatorname{KL}(P,Q)=\operatorname{KL}(p,q) = \int {p(y)} \log \frac{p(y)}{q(y)} \mathrm{d} y.
\]
The Kullback-Leibler divergence has the natural property of being nonnegative, and it equals zero if and only if $P=Q$ (equivalently, when $p=q$, almost everywhere). Many other useful properties of the Kullback-Leibler divergence are collected in \cite{cover06}. The Kullback-Leibler divergence admits a fundamental information-theoretic interpretation, in that $\operatorname{KL}(P,Q)$ can be interpreted as the `surprise' experienced on average when one believes that $Q$ describes a given probabilistic phenomenon and is then told that it is in fact described by $P$ (\cite{white94}, p. 9). The Kullback-Leibler divergence admits a straightforward generalisation to conditional distributions.

For each competing model, under mild regularity conditions there will be a parameter value $\theta^{\ast},$ that minimises the Kullback-Leibler divergence between the conditional probability densities $p_0(\cdot | \cdot)$ and $p(\cdot | \cdot; {\theta})$, i.e.\
\begin{equation}
\label{kl}
\mathbb{E}_0 \left[ \log \frac{p_0(Y_i|t_i)}{p(Y_i | t_i;\theta)} \right].
\end{equation}
Here the expectation is under the true joint distribution $P_0$ of the pair $(t_i,Y_i).$ Thus the density $p(\cdot | \cdot; {\theta^{\ast}})$ constitutes the best approximation to $p_0(\cdot | \cdot)$ among the densities $p(\cdot | \cdot; {\theta}).$ The parameter value $\theta^{\ast}$ is referred to as the pseudo-true value of $\theta$, while $p(\cdot | \cdot; {\theta^{\ast}})$ is called the pseudo-true model (see \cite{sawa78}, p.~1276 and \cite{vuong89}, p.~308).

Minimisation of Equation \eqref{kl} over $\theta$ is equivalent to maximisation of
\begin{equation}
\label{kl2}
\mathbb{E}_0 \left[ p(Y_i | t_i;\theta) \right],
\end{equation}
over $\theta$ and yields the pseudo-true value $\theta^{\ast}$; note that this latter implicitly depends on $P_0.$ Unfortunately, since \eqref{kl2} depends on the unknown distribution $P_0$, the Kullback-Leibler divergence is not computable in practice and thus neither can be minimized over $\theta.$ However, it can be estimated by the sample average
\[
\frac{1}{n}\sum_{i=1}^n \log p(Y_i | t_i;\theta),
\]
which can be maximized instead  (see \cite{white94}, Section 2.3). Equivalently, one can maximize $\sum_{i=1}^n \log p(Y_i | t_i;\theta).$ This amounts to nothing else but employing the well-known maximum likelihood method devised by R.~A.\ Fisher as a general parameter estimation technique in statistical problems; see \cite{fisher22} and \cite{fisher25}. The latter is a popular and, under rather general assumptions, statistically optimal approach to parameter estimation; see, e.g., \cite{vandervaart98} and \cite{white94} for modern accounts of the theory. The (conditional) maximum likelihood estimator (MLE) of the parameter $\theta$ is defined as
\begin{equation}
\label{mle}
\hat{\theta}_n=\operatorname{argmax}_{\theta\in\Theta} \sum_{i=1}^n \log p(Y_i | t_i;\theta).
\end{equation}
Under rather general conditions, that we do not list here and refer to e.g.\ \cite{white94}, Chapters 3--7 instead, the MLE $\hat{\theta}_n$ exists, converges to the pseudo-true value $\theta^{\ast}$, and is in fact asymptotically normal and optimal in a specific sense. Thus the maximum likelihood estimator $\hat{\theta}_n$ can be used as a proxy for the pseudo-true value $\theta^{\ast}$, and consequently the conditional density $p(\cdot | \cdot;\hat{\theta}_n)$ can serve as a proxy for $p(\cdot | \cdot ; \theta^{\ast})$, and eventually for the true conditional density $p_0(\cdot | \cdot).$
\subsection{Testing Framework}
\label{subsect:discrimination}

\begin{figure}
\centering
\includegraphics[width=11cm]
{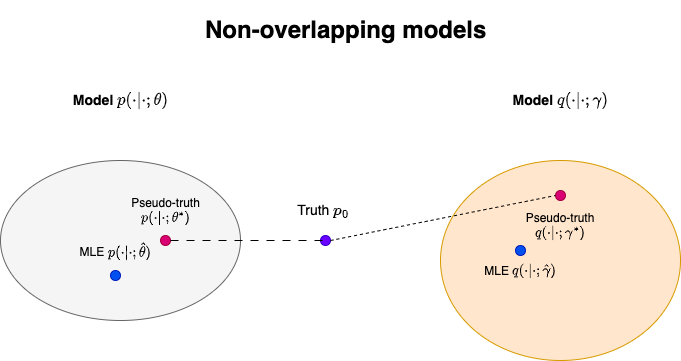}
\caption{Schematic depiction of two non-overlapping statistical models.}
\label{fig:models}
\end{figure}

Suppose next to $p(\cdot | \cdot ; \theta)$ we have another family of conditional densities $q(\cdot | \cdot ; \gamma),$ parametrised by $\gamma\in\Gamma$ (defined according to Equation \ref{theta}), as a possible approximation to $p_0(\cdot | \cdot).$ We denote by $\gamma^{\ast}$ the pseudo-true value corresponding to the family $q(\cdot | \cdot ; \gamma)$ and by $\hat{\gamma}_n$ the maximum likelihood estimator. \cite{vuong89} considered the following formal framework for model discrimination in this context:
\[
H_0: \mathbb{E}_0 \left[ \log p(Y_i | t_i;\theta^{\ast}) \right] = \mathbb{E}_0 \left[ \log q(Y_i | t_i;\gamma^{\ast}) \right],
\]
meaning the models $p(\cdot | \cdot ; \theta)$ and $q(\cdot | \cdot ; \gamma)$ are equivalent, versus an alternative hypothesis
\[
H_p: \mathbb{E}_0 \left[ \log p(Y_i | t_i;\theta^{\ast}) \right] > \mathbb{E}_0 \left[ \log q(Y_i | t_i;\gamma^{\ast}) \right],
\]
meaning the model $p(\cdot | \cdot ; \theta)$ is better than the model $q(\cdot | \cdot ; \gamma)$, and another alternative
\[
H_q: \mathbb{E}_0 \left[ \log p(Y_i | t_i;\theta^{\ast}) \right] < \mathbb{E}_0 \left[ \log q(Y_i | t_i;\gamma^{\ast}) \right],
\]
meaning the model $q(\cdot | \cdot ; \gamma)$ is better than the model $p(\cdot | \cdot ; \theta)$. The choice between $p(\cdot | \cdot ; \theta)$ and $q(\cdot | \cdot ; \gamma)$ is called model testing. It results in selection of one model as the best (in the sense that it is closer to the $p_0(\cdot | \cdot)$), or retaining the null hypothesis that both models are equally accurate (or inaccurate). The reader may find Figure \ref{fig:models} and other similar ones helpful when trying to visualise schematically various quantities and concepts mentioned throughout this section.

The above-displayed formulae suggest a natural way to proceed with testing: replace the information quantities with their sample analogues and base a decision on the log-likelihood ratio
\begin{equation}
\label{lr}
\operatorname{LR}_n=\sum_{i=1}^n \log \frac{ p(Y_i | t_i;\hat{\theta}_n) }{ q(Y_i | t_i;\hat{\gamma}_n) }.
\end{equation}
This latter (or equivalently, the likelihood ratio) has been used extensively and with great success in various testing problems (see, e.g., \cite{vandervaart98}, Chapter 16 for a modern account). Vuong derived the asymptotic distribution of \eqref{lr} under $H_0$, as well as its limits under $H_p$ and $H_q,$ thereby obtaining critical values, and also concluding that the test is directional: the rejection of the null $H_0$ occurs in the direction of either $H_p$ or $H_q.$

Unfortunately, the limiting distribution of the log-likelihood ratio in Vuong's framework depends in a complicated way on the relationship between the models $p(\cdot | \cdot ; \theta)$ and $q(\cdot | \cdot ; \gamma)$, that is not easy to ascertain for ODE models. The key difficulty is nestedness of the models. The two models
\[
P_{\theta}=\{ p(\cdot | \cdot ; \theta): \theta \in \Theta \}, \quad Q_{\gamma}=\{ p(\cdot | \cdot ; \gamma): \gamma \in \Gamma \}
\]
are said to be strictly non-nested, if $P_{\theta} \cap Q_{\gamma} = \varnothing.$ They are said to be nested, if either $P_{\theta} \subset Q_{\gamma}$ or $P_{\theta} \supset Q_{\gamma}.$ Finally, they are said to be overlapping, if  $P_{\theta} \cap Q_{\gamma} \neq \varnothing$ and either $P_{\theta} \not\subset Q_{\gamma}$ or $P_{\theta} \not\supset Q_{\gamma}$, or both. See \cite{vuong89} and cf.\ \cite{pesaran87} and \cite{mcaleer_pesaran86} for additional details and examples. For ODE-based statistical systems, verification of the relationship between models reduces to verification of the relationship between the ODE systems. Since these typically do not admit closed-form solutions, the exact relationship between the two ODE models will remain unclear.
As such, in many if not most cases, Vuong's test will necessarily involve a pre-test step, and the resulting two-stage test may exhibit a significant size distortion. The latter is a serious defect and its extent has been demonstrated in \cite{shi15}, Section 3. Methods to address this issue have been proposed by \cite{schennach17} and \cite{shi15}. The former is arguably simpler to present and implement and moreover works without modification irrespective of the relationship between the models $p(\cdot | \cdot ; \theta)$ and $q(\cdot | \cdot ; \gamma)$. Hence our decision to concentrate on the Schennach-Wilhelm test in this research.

\section{Methodological Approach}
\label{subsect:method}
\subsection{Schennach-Wilhelm test}
\label{subsect:schennach}

Let $h_n>0$ denote a data-dependent regularisation parameter; we will discuss its choice below. Assume for simplicity the number of observations $n$ is even, and introduce the weights
\[
w_k(h_n)=
\begin{cases}
1, & k \textrm{ odd} ,\\
1+h_n, & k \textrm{ even},
\end{cases}
\quad k=1,\ldots,n+1.
\]
Define the reweighted log-likelihood ratio
\[
\widetilde{\operatorname{LR}}_n=\frac{1}{n}\sum_{i=1}^n \left( w_i(h_n) \log { p(Y_i | t_i;\hat{\theta}_n) } - w_{i+1}(h_n) \log { q(Y_i | t_i;\hat{\gamma}_n) } \right),
\]
and let (see Supplementary Material for the derivation)
\[
\hat{\widetilde{\sigma}}^2=(1+h_n)\hat{\sigma}^2+\frac{h_n^2}{2}(\hat{\sigma}_p^2+\hat{\sigma}_q^2)
\]
be an estimator of the asymptotic variance of the reweighted log-likelihood ratio. Here
\[
\hat{\sigma}^2=\hat{\sigma}_p^2-2\hat{\sigma}_{pq}+\hat{\sigma}_q^2
\]
with
\begin{align*}
\hat{\sigma}_p^2 & = \frac{1}{n}\sum_{i=1}^n \left( \log  p(Y_i | t_i;\hat{\theta}_n) - \overline{\log p} \right)^2,\\
\hat{\sigma}_{pq} & = \frac{1}{n}\sum_{i=1}^n \left( \log  p(Y_i | t_i;\hat{\theta}_n) - \overline{\log p} \right) \left( \log  q(Y_i | t_i;\hat{\gamma}_n) - \overline{\log q} \right),  \\
\hat{\sigma}_q^2 & = \frac{1}{n}\sum_{i=1}^n \left( \log  q(Y_i | t_i;\hat{\gamma}_n) - \overline{\log q} \right)^2,
\end{align*}
and we have used the notation
\[
\overline{\log p} = \frac{1}{n} \sum_{i=1}^n \log  p(Y_i | t_i;\hat{\theta}_n), \quad \overline{\log q} = \frac{1}{n} \sum_{i=1}^n \log  q(Y_i | t_i;\hat{\gamma}_n).
\]
The Schennach-Wilhelm test statistic is defined as
\[
\widetilde{T}_n=\frac{\sqrt{n}\widetilde{LR}_n}{ \hat{\widetilde{\sigma}} }.
\]
It is shown in \cite{schennach17}, Theorem 1, that under regularity assumptions on the statistical models under consideration, that we do not list here but refer to the original paper, the statistic $\widetilde{T}_n$ is asymptotically normal under $H_0$, i.e. $\widetilde{T}_n\sim N(0,1),$ diverges to $+\infty$ under $H_p,$ and to $-\infty$ under $H_q.$ This asymptotic result readily yields a test for model selection. Namely, fix a level $0<\alpha<1$. Let $z_{1-\alpha/2}$ be the $1-\alpha/2$-quantile of the standard normal distribution. Then retain $H_0$ if $|\widetilde{T}_n|\leq z_{1-\alpha/2},$ and otherwise reject it. Rejection occurs in favour of $H_p,$ if $\widetilde{T}_n > z_{1-\alpha/2},$ and in favour of $H_q,$ if $\widetilde{T}_n < - z_{1-\alpha/2}.$ \cite{schennach17}, Section 5, establish favourable theoretical properties of their test. They also conduct a simulation study and apply their method on a real data set. However, the practical examples they consider are limited to classical statistical models like normal location and linear regression. Application of the Schennach-Wilhelm test to ODE models is a novel contribution.

\subsection{Regularisation Parameter}
\label{sect:hn}
In this subsection, following \cite{schennach17}, we present a methodology for an optimal choice of the regularisation parameter $h_n.$ This involves a step that is specific only to ODE-based models. Obviously, the parameter $h_n>0$ can also be chosen subjectively, according to a researcher's preferences, but \cite{schennach17} present the following objective methodology: they remark that their test has a size distortion only if the two models are nested or overlapping. This can be controlled by choosing the regularisation parameter $h_n$ large. On the other hand, taking $h_n$ large makes the test lose power in the case of strictly non-nested models. Schennach and Wilhelm suggest choosing $h_n$ that balances the worst cases under these two scenarios. They achieve this by expanding the test size and power in these two cases in terms of $h_n$, and next balancing the terms in the two expansions by choosing an appropriate $h_n$ (and estimating some constants from the data). Instead of reporting the details of such computations here, that can be found \cite{schennach17}, Section 6, we directly provide the formula for the optimal $\hat{h}_n$ they derived,
\[
\hat{h}_n=\left( \frac{\hat{C}_{SD}}{\hat{C}_{PL}} \right)^{1/3} n^{-1/6}( \log\log n )^{1/3}.
\]
Here the two constants $\hat{C}_{SD},\hat{C}_{PL}$ are
\begin{align*}
\hat{C}_{SD}&=\phi\left( z_{\alpha/2} - \frac{\hat{\delta}}{\hat{\sigma}} \right) \frac{ \hat{\delta} (\hat{\sigma}^2 - 2 (\hat{\sigma}_p^2+\hat{\sigma}_q^2)) }{4\hat{\sigma}^3},\\
\hat{C}_{PL}&=2\phi(z_{\alpha/2}) \frac{ \max \{ |\operatorname{tr}(\hat{H}_p^{-1}\hat{V}_p)|,  |\operatorname{tr}(\hat{H}_q^{-1}\hat{V}_q)| \} }{ \sqrt{(\hat{\sigma}_p^2+\hat{\sigma}_q^2)/2} },
\end{align*}
and we used the notation
\begin{align*}
\hat{\delta}&=\frac{\hat{\sigma}}{2}(z_{\alpha/2}-\sqrt{4+z_{\alpha/2}^2}),\\
\hat{H}_p&=\frac{1}{n}\sum_{i=1}^n \nabla_{\theta}^2\log p(Y_i|t_i;\hat{\theta}_n),\\
\hat{V}_p&=\frac{1}{n}\sum_{i=1}^n (\nabla_{\theta}\log p(Y_i|t_i;\hat{\theta}_n) (\nabla_{\theta}\log p(Y_i|t_i;\hat{\theta}_n)^{\prime} ,
\end{align*}
and similarly for $\hat{H}_q$ and $\hat{V}_q$. The operator $\nabla_{\theta}$ gives the gradient with respect to $\theta,$ while $\nabla_{\theta}^2$ gives the Hessian. The matrices $\hat{H}_p,\hat{V}_p$ (as well as those corresponding to the conditional density $q(\cdot|\cdot;\gamma)$) are needed for the usual sandwich variance estimator in potentially misspecified models (see, e.g., \cite{white94}, Section 8.3), and hence the requirement for their computation does not go beyond what is done when using the maximum likelihood method (see \cite{schennach17}, Section 6). On the other hand, evaluation of the derivatives $\nabla_{\theta}\log p(Y_i|t_i;\hat{\theta}_n)$ and $\nabla_{\theta}^2\log p(Y_i|t_i;\hat{\theta}_n)$ has some peculiarities in the ODE context, but nevertheless is conceptually straightforward, in that it reduces to numerical integration of the sensitivity and variational equations associated with the ODE system. Here we concentrate on the case when the ODE systems under consideration are one-dimensional for simplicity of exposition. A generalization of the arguments to the multidimensional case is straightforward, but much more involved notationally (see Supplementary Material for details).

What we are interested in are the derivatives $\nabla_{\theta} \log p(y|t;\theta)$ and $\nabla_{\theta}^2 \log p(y|t;\theta).$ We aggregate the pair $\xi,\psi$ into a vector $\eta=(\xi,\psi)$, where $\psi$ is the vector of ODE system parameters, and $\xi$ is the vector of the state initial values. At this stage we need to make concrete the distributional assumptions on the likelihood, and we assume it is Gaussian. Nevertheless, the same roadmap can be adapted  to alternative likelihood functions as well. Under Gaussianity,
\[
\log p(y|t;\theta)=-\frac{1}{2}\log(2\pi\sigma^2)-\frac{(y-x(t;\eta))^2}{2\sigma^2}.
\]
Calculating the first and second-order partial derivatives with respect to $\sigma^2$ is  straightforward. Here we concentrate on derivatives with respect to $\eta.$ This boils down to evaluating the derivatives
\[
\nabla_{\eta} (y-x(t;\eta))^2, \quad \nabla_{\eta}^2 (y-x(t;\eta))^2.
\]
Now
\[
\nabla_{\eta} (y-x(t;\eta))^2=-2 (y-x(t;\eta)) \times \nabla_{\eta} x(t;\eta).
\]
We should thus find means for computing $\nabla_{\eta} x(\cdot;\eta).$ This is, however, standard via numerical integration. Differentiate both sides of Equation \eqref{eq:ode} with respect to $\eta$, interchange on the lefthand side the order of the $\eta$- and $t$-derivatives, define $s(t)=\nabla_{\eta}{x}(t;\eta)$, and get
\begin{equation}
\label{nomer2}
\begin{split}
\frac{\mathrm{d} s}{\mathrm{d} t}&=\frac{\partial}{\partial x} F({ x}({\eta},t),{\eta})s(t)+\frac{\partial}{\partial\eta} F(x(t;{\eta}),{\eta}),\\
s(0)&=(1,0)^{\prime},
\end{split}
\end{equation}
Here $1$ and $0$ in the initial condition are understood as vectors of $1'$s and $0'$s of appropriate dimensions, and the notation $\frac{\partial}{\partial x} F$ and $\frac{\partial}{\partial\eta} F$ stands for the derivatives of the function $F$ with respect to its first and second arguments, $x$ and $\eta$. As $x$ is a known function from Equations \eqref{eq:ode}, \eqref{nomer2} is a linear system with known time-dependent coefficients, and hence is relatively easy to integrate. It is referred to as the system of sensitivity equations. Some care needs to be taken when integrating Equation \eqref{nomer2}, as it is typically a stiff system, but by now numerical integration techniques for such systems have been well-studied; see, e.g., \cite{hairer10}.

In a similar manner, setting $z(t)=\nabla_{\eta}^2 x(t;\eta),$ we can get the matrix differential equation, called the system of variational equations,
\begin{equation}
\label{nomer3}
\begin{split}
\frac{dz}{dt}&=\frac{\partial^2}{\partial\eta^2}F( x(t;{\eta}),{\eta} )+\frac{\partial^2}{\partial\eta\partial x}F(x(t;{\eta}),{\eta}) s(t)\\
&+\left\{ \frac{\partial^2}{\partial\eta\partial x }F (x(t;{\eta}),{\eta}) + \frac{\partial^2}{\partial x^2}F (x(t;{\eta}),{\eta}) s(t) \right\}s(t)\\
&+\frac{\partial}{\partial x}F( x(t;{\eta}),{\eta} )z(t),\\
z(0)&=0,
\end{split}
\end{equation}
where the initial condition is a zero matrix of appropriate dimensions, and partial derivatives are derivatives of the function $F$ with respect to its first and/or second arguments. Again, $x$ and $s$ in Equation \eqref{nomer3} are known, and the system is linear with time-dependent coefficients. Although the system might be stiff, its integration is a well-studied task.

\subsection{Algorithm }
\label{subsec:mlalgo}

With all the technical steps worked out, our model selection procedure is summarised as follows:

\begin{enumerate}
\item Estimate parameters $\eta = (\xi, \psi)$ for each model using    
$\hat{\eta}_n = \operatorname{argmin}_{\eta}\sum_{j=1}^d\sum_{i=1}^n(Y_{ji}-x_j(t_i;\eta))^2.$
\item Estimate parameters $\sigma^2$ for each model using $\hat{\sigma}_{j}^2 = \frac{1}{n}\sum_{i=1}^n(Y_{ji}-x_j(t_i;\hat{\eta}))^2.$ 
\item Obtain numerical derivatives ($\hat{V}_n$ and $\hat{H}_n$ matrices) according to Equations \eqref{nomer2} and \eqref{nomer3}, or more generally the ones in the Supplementary Material.
\item Create $\frac{N^2-N}{2}$ model pairs to test, where N is the total number of competing models.
\item Calculate the regularization parameter according to Section \ref{sect:hn}, using matrices $\hat{V}_n$, $\hat{H}_n$.
\item Calculate the S-W test statistic for each pair of models according to Section \ref{subsect:schennach}.
\item Compare the outcomes of the S-W test statistic to the critical values from the standard normal distribution.
\end{enumerate}

\section{Simulation Study: S-W Test for ODEs}\label{sec:odetypes}

In this section, we present results of simulations aiming to show the size and power properties of the S-W test when applied to the ODE model selection. For asymptotic properties of the test we refer the reader to Section 5 in \cite{schennach17}.
\label{sims}

 \begin{figure}
     \centering
     \begin{subfigure}[b]{0.48\textwidth}
         \centering
         \includegraphics[width=\textwidth]{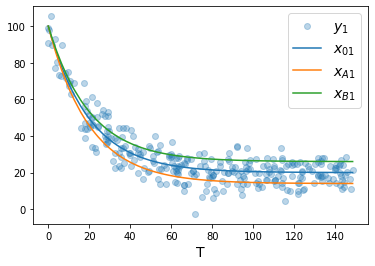}
        \caption{DGP and models fits when $\delta=0.3$}\hfill
         \label{fig:size_sim_curves}
     \end{subfigure}
     \hfill
 \begin{subfigure}[b]{0.48\textwidth}
     \centering
     \includegraphics[width=\textwidth]{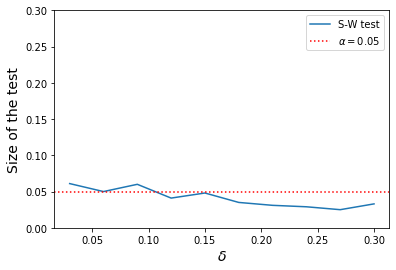}
    \caption{The size of the test depending on the value of $\delta$}\hfill
     \label{fig:size_sim_res}
 \end{subfigure}
 \hfill
 \begin{subfigure}[b]{0.48\textwidth}
     \centering         \includegraphics[width=\textwidth]{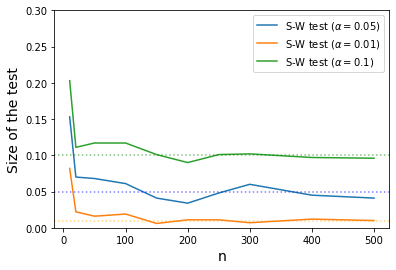}
     \caption{Size of the test for different levels of $\alpha$}\hfill
     \label{fig:size_sim_asymp}
 \end{subfigure}
 \hfill
 \begin{subfigure}[b]{0.48\textwidth}
     \centering         \includegraphics[width=\textwidth]{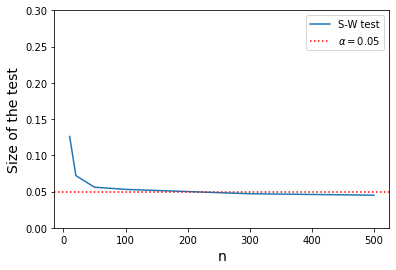}
     \caption{Size of the test on the equally distant time grid}\hfill
     \label{fig:size_eq_dist}
 \end{subfigure}
 \hfill
    \caption{Plots describing the model setups and results of the test size simulations.}
    \label{fig:size_sim_imgs}
\end{figure}

\subsection{Size of the Test}
\label{test_size}
Nominally the test size is  $\alpha$, where $\alpha$ is chosen by the user. However, due to various approximations involved, in practice the size may significantly deviate from the nominal level. This is an undesirable behaviour, but its extent cannot be fully studied through theoretical means only. In order to assess whether the test controls the size at the desired level, we designed a simulation study. The details are as follows: let the data generation process (DGP) correspond to the ODE system 
\[
\begin{cases}
x_0^{\prime}(t) & =-0.05x_0(t) + 1,\\
x_0(0) & = 100,\\
\end{cases}
\]
and observations be sampled as
\[
y(t_i) = x_0(t_i) + \epsilon_i
\]
where $\epsilon\sim N(0, 7^2)$. Define model A (left) and model B (right):
\begin{center}
\begin{tabular}{ c c }   
$\left\{\begin{matrix}
x_A^{\prime}(t) = -0.05x_A(t) + (1-\delta)\\
x_A(0) = \xi_A \hfill
\end{matrix}\right.$ 
 &
$\left\{\begin{matrix}
x_B^{\prime}(t) = -0.05x_B(t) + (1+\delta)\\
x_B(0) = \xi_B \hfill
\end{matrix}\right.$ 
\end{tabular}
\end{center}
Here $\delta$ is a known constant that shifts models from the DGP. The only unknown parameters are the initial values $\xi$. As an illustration, Figure \ref{fig:size_sim_curves} shows simulated observations, as well as the curves of the truth and A and B models. Here  $\delta=0.3, \xi_{0}=\xi_{A}=\xi_{B}=100,\psi_{0}=\psi_{A}=\psi_{B}=-0.05$, number of observations is 300 and the range of time points is up to $\tau=150$. Both models and the DGP can be written as
\[
\left\{\begin{matrix}
x_0^{\prime}(t) =\psi_1x_0(t) + \psi_2\\
x_0(0) = \xi_0 \hfill
\end{matrix}\right.
\]
The solution to this nonhomogeneous linear ODE is
\[
x_0(t) = \xi_0 e^{\psi_1 t} + \frac{\psi_2}{\psi_1} e^{\psi_1 t} - \frac{\psi_2}{\psi_1}.
\]
The DGP corresponds to $\psi_2 = 1$, while Models $A$ and $B$ are determined, respectively, by $\psi_2 = 1 - \delta$ and $\psi_2 = 1 + \delta$. With Gaussian errors, and the assumption that the errors are independet of observation times, the Kullback-Leibler divergence between the DGP and either Model A or B is
\[
\frac{1}{2\sigma^2}\frac{\delta^2}{\psi_1^2} \int_0^T \left(1 - e^{\psi_1 t}  \right)^2 f_T(t) dt,
\]
where $f_T(\cdot)$ is the density of $t_i$'s with support on $[0,T]$. The factor 
\[
\frac{1}{2\sigma^2}\frac{1}{\psi_1^2} \int_0^T \left(1 - e^{\psi_1 t}  \right)^2 f_T(t) dt
\]
is a constant, and hence the divergence as a function of $\delta$ scales as $\delta^2$, meaning that whenever $\delta$ is the same in Models A and B, the KL-divergences between them and the DGP are equal.

To calculate the real level at which the test controls size (the probability to reject $H_0$ when it is true), we ran 1000 simulations for $\delta=0.03, 0.06, \ldots, 0.3$ and $\alpha=0.05$. The results of this experiment can be observed in Figure \ref{fig:size_sim_res}: as we see the test controls the size around the desired level $\alpha=0.05$ for any $\delta$. Furthermore, Figure \ref{fig:size_sim_asymp} shows that the S-W test asymptotically controls the size for any level of alpha (the simulation setup is the same as above with $\delta=0.1$ and $n\in [10, 20, 50, 100, 150, 200, 250, 300, 400, 500]$). Additionally, we provide results for the size of the test run on an equally distant time grid to show that deterministic time in our approach works similarly to uniform one (compare Figure \ref{fig:size_eq_dist} to Figure \ref{fig:size_sim_asymp} with $\alpha=0.05$).

\subsection{Power of the Test}
\label{test_power}
We conducted two sets of simulations to assess the power of the S-W test. We considered from Section \ref{subsect:real_exam2} Model 1 and Model 2 reparameterized as follows:
\begin{align*}
x_1^{\prime}(t) &= \psi_2\psi_3x_1(t)x_2(t)-\psi_4x_1(t),\\
x_2^{\prime}(t) &=\psi_1x_2(t)\left(1-\psi_5x_2(t)\right)-\psi_2x_1(t)x_2(t),\\
\textbf{x(0)} & = [\xi_1, \xi_2];
\end{align*}
Hence the old parameter $\psi_5$ became $1/\psi_5$ in the new parametrization. The motivation here is that now with $\psi_5=0$ the models are equivalent, while with $\psi_5 > 0$ they are not. In the old parametrization, equivalence would require $\psi_5\rightarrow \infty$, which would numerically destabilize estimation.

\begin{figure}
     \centering
     \begin{subfigure}[b]{0.48\textwidth}
         \centering        
         \includegraphics[width=\textwidth]{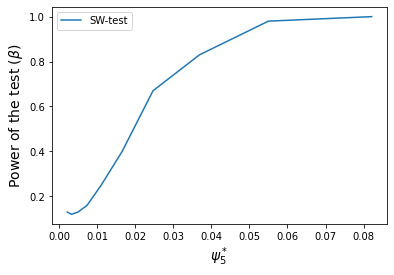}
         \caption{Power of the test ($n=20$)}   
         \label{fig:power_n40}
     \end{subfigure}
     \hfill
     \begin{subfigure}[b]{0.48\textwidth}
         \centering
         \includegraphics[width=\textwidth]{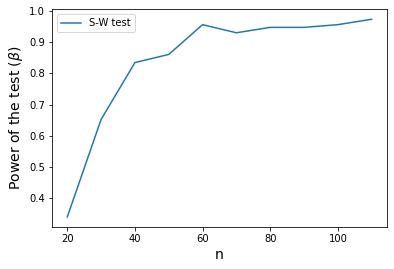}
         \caption{Power of the test ($\psi_5=0.1$)}
         \label{fig:power_over_n}
     \end{subfigure}
        \caption{Simulated power of the S-W test. For details, see the text in Section \ref{test_power}.}
        \label{fig:swtest_power}
\end{figure}

For the first set of simulations, we set Model 2 as the DGP with $\theta=(\sigma_1^2=0.1^2, \xi=[1,2],\psi=[1,1,1,1,\psi_5])$, where $\psi_5$ varies between 0.0025 and 0.25. The number of time points is 20, $\tau=40$ and the number of simulations per each $\psi_5$ is 100. The simulated power of the test is then presented in Figure \ref{fig:power_n40}. 

For the second set of simulations, the DGP is Model 2 with $\theta=(\sigma_1^2=0.2^2, \xi=[1,2],\psi=[1,1,1,1,0.1])$ and we let the sample size $n$ vary from 20 to 110 with steps of 10. Figure \ref{fig:power_over_n} shows the simulated power of the test.

From the two figures, it can be concluded that the test works as expected, because:
\begin{itemize}
    \item with a small number of observations, discrimination between models requires larger KL-distance between them and the truth (Figure  \ref{fig:power_n40});
    \item when both models are close to DGP, a larger number of observations is required to reveal the model that is closer to the truth (Figure  \ref{fig:power_over_n}).
\end{itemize}

\section{Discussion and Outlook}
\label{discussion_outlook}
\subsection{Importance of Assumptions}
In situations involving uncertainty, optimal solutions are often elusive. The S-W test, as in fact any other statistical test, is derived under specific theoretical assumptions, and it is imperative to bear them in mind when applying the test in practice. Among these, the most significant assumption is that of independent and identically distributed observations, as well as various regularity conditions on the statistical model. A researcher, in line with advice in \cite{snedecor1989}, p.~273, must either possess confidence or take supplementary steps to verify assumptions when utilizing the S-W test for ODE model selection.

\subsection{From Theory to Practice}
Another important consideration is the step from a theoretical or algorithmic description of an estimation or testing method to its implementation and performance in practice. As well-documented in the literature, statistical inference in ODE models presents unique numerical challenges. Hence also when using the S-W test, the user must ensure that optimization routines and numerical integrators perform as required and do not invalidate the inferential conclusions due to a failure. Here the best practices as described, e.g., in \cite{ramsay2017dynamic}, but also in general nonlinear regression literature, and, above all, practical experience with fitting ODE models, are most helpful.

\subsection{Desirable Extensions}

Although \cite{schennach17} mention a potential extension of their method to the time series data, they do not furnish a comprehensive theoretical validation for this claim. The details can be worked out, however. This would take care of issues like correlated errors, that arise in estimation problems for ODEs. What is less clear is a possibility of extension of the S-W test to the mixed model setting (see, e.g., \cite{pinheiro2000}), which would be of great practical relevance.

\subsection{Outlook}

Modeling with ordinary differential equations (ODEs) serves as an indispensable tool across a myriad of scientific and engineering domains. From mathematical biology to compartmental models in epidemiology, ODEs offer a comprehensive understanding of intricate dynamics. Grounded in the foundational laws of nature, these mechanistic models aim at shedding light on the causal underpinnings of systems, proving invaluable especially when direct experimentation is constrained by feasibility or cost considerations.

However, the ODE landscape is intricate. Often, a single phenomenon can be expressed through multiple models, originating from approximations of internal mechanisms or from variances in the abstraction level. When faced with this plurality, the quandary arises: how do we select the best-fitting model amidst the obscurities of statistical noise?

Rather than delving into deterministic model discrimination, our study champions a testing-based approach. Anchored in the model misspecification framework, our focus narrows to identifying the most apt explanation for the data-generating mechanism within a given parametric model. Drawing from the foundational insights of Vuong and Hotelling, our methodology sidesteps the oft-tedious nesting requirements, streamlining the comparison and ranking of distinct causal explanations. However, the allure of a two-step testing mechanism is not without its pitfalls. Schennach and Wilhelm's critiques highlight potential distortions. Their modification, the S-W test, stands as a solution, offering robustness irrespective of nestedness of the statistical models. While Vuong's and Schennach \& Wilhelm's paradigms were primarily designed for traditional statistical frameworks, our research has tailored them to the ODE context. This adaptation, though intricate, has been addressed with meticulous attention to detail.

To fortify the claimed theoretical stance, we undertook rigorous simulation studies. These were instrumental in illustrating both the size and power of our proposed test, closely aligning with theoretical predictions. Moreover, through a series of real data examples, we showcased the practical applicability of our algorithm, underscoring its versatility and adaptability across various scenarios. Beyond the realm of theory and experimentation, we believe in the democratization of knowledge. Recognizing the importance of accessibility and hands-on application, we've provided a user-friendly Python implementation of our model selection algorithm (\href{https://github.com/odemodelselection/sw_test}{S-W test on GitHub}). This not only fosters a deeper understanding but also empowers researchers and practitioners to implement our findings directly, paving the way for further innovations in the field. 

To conclude, as the mathematical modeling landscape continues to evolve, our findings and contributions seek to continually refine and advance ODE model selection methodologies for the broader scientific community.

\bibliography{main}

\end{document}


\title{Supplementary Material for Model Selection for Ordinary Differential Equations: a Statistical Testing Approach}

\author{Itai Dattner\orcidA{}
\\Department of Statistics, University of Haifa
\\idattner@stat.haifa.ac.il
\and
Shota Gugushvili\orcidB{}
\\Biometris, Wageningen University \& Research
\\shota.gugushvili@wur.nl
\and
Oleksandr Laskorunskyi\orcidA{}
\\Department of Statistics, University of Haifa
\\olaskoru@campus.haifa.ac.il

\maketitle

\section{Recent Model Selection Approaches}
\label{recentms}
\begin{center}
\small
    \centering
    \begin{tabular}{||p{0.2in} p{4in} p{2in} ||} 
    \hline
     Year & Criterion & Reference\\ [0.5ex] 
    \hline\hline
2022	& Stability-based model selection method PDE-STRIDE	& \cite{maddu2022stability}\\
2021	& Weak SINDy	& \cite{messenger2021weak}\\
2020	& ODIN: ODE-Informed Regression	& \cite{wenk2020odin} \\
2020	& Accumulated Prediction Error (APE)	& \cite{ippolito2020studying}\\
2020	& Symmetry-based methods to a model selection& \cite{ohlsson2020symmetry}\\
2020    & Predicted derivative	& \cite{kaheman2020sindy}\\
2019	& Marginal likelihood estimation by combining thermodynamic integration and gradient matching	& \cite{macdonald2019model}\\
2016	& Occam Plausibility Algorithm (OPAL)	&\cite{lima2016selection}\\
2015	&  LSA criterion&\cite{zhang15}\\
2015	& Stratified random cross-validation (SRCV) approach&\cite{hasdemir15validation}\\
2013	& General mixed-integer nonlinear programming (MINLP) formulation	& \cite{rodriguez2013simultaneous}\\
2012& 	Prediction discrepancies (pd) and normalised prediction distribution errors (npde)	& \cite{nguyen2012extension}\\
2011	& A least angle regression (LARS) method	& \cite{li2011model}\\
2011& 	ABC SMC	& \cite{toni2011abc}\\
2008	& SEM algorithm&\cite{liu2008gene}\\
    \hline
    \end{tabular}
\end{center}

\section{Variance of $\sqrt{n}\hat{\Tilde{d}}$}
\label{sigmatildahat}
We need to show that the asymptotic variance of $\sqrt{n}\hat{\Tilde{d}}$ is
\[
\hat{\Tilde{\sigma}}^2 = (1 + h_n)\hat{\sigma}^2 + \frac{h^2_n}{2}(\hat{\sigma}^2_A + \hat{\sigma}^2_B)
\]
We are given that
\[
\hat{\Tilde{d}} = \frac{1}{n}\sum_{i=1}^n(w_i(h_n)logf_A(X_i;\theta_A)-w_{i+1}(h_n)logf_B(X_i;\theta_B))
\]
\[
w_k(h_n)=
\left\{\begin{matrix}
1, if\ k \ is \ odd\\
1+h_n, if\ k \ is \ even
\end{matrix}\right.
\]
Define $A_i=\log f_A(X_i;\theta_A), B_i = \log f_B(X_i;\theta_B)$ and consider the case $n=2$. Then
\[
\hat{\Tilde{d}} = \frac{1}{2}(A_1-(1+h_n)B_1 + (1+h_n)A_2-B_2)= 
\]
\[
VAR[\hat{\Tilde{d}}] = \frac{1}{4}VAR[(A_1-(1+h_n)B_1] + \frac{1}{4}VAR[(1+h_n)A_2-B_2)]= \]
\[
= \frac{1}{4}[\hat{\sigma}^2_A-2(1+h_n)\hat{\sigma}_{AB}+(1+h_n)^2\hat{\sigma}^2_B + (1+h_n)^2\hat{\sigma}^2_A-2(1+h_n)\hat{\sigma}_{AB}+\hat{\sigma}^2_B)]= \]
\[
= \frac{1}{4}[\hat{\sigma}^2_A-4\hat{\sigma}_{AB}-4h_n\hat{\sigma}_{AB}+\hat{\sigma}^2_B+2h_n\hat{\sigma}^2_B+h_n^2\hat{\sigma}^2_B + \hat{\sigma}^2_A+2h_n\hat{\sigma}^2_A+h_n^2\hat{\sigma}^2_A+\hat{\sigma}^2_B]= \]
\[
= \frac{1}{4}[2\hat{\sigma}^2_A +2h_n\hat{\sigma}^2_A-4\hat{\sigma}_{AB}-4h_n\hat{\sigma}_{AB}+2\hat{\sigma}^2_B+2h_n\hat{\sigma}^2_B+h_n^2\hat{\sigma}^2_B+h_n^2\hat{\sigma}^2_A]= \]

\[
= \frac{1}{4}[2(1 +h_n)\hat{\sigma}^2_A-4(1+h_n)\hat{\sigma}_{AB}+2(1+h_n)\hat{\sigma}^2_B+h_n^2\hat{\sigma}^2_B+h_n^2\hat{\sigma}^2_A]= \]
\[
= \frac{1}{4}[2(1 +h_n)\hat{\sigma}^2+h_n^2(\hat{\sigma}^2_B+\hat{\sigma}^2_A)]
\]
In the general case we get
\[
VAR[\hat{\Tilde{d}}] = \frac{1}{n^2}[n(1 +h_n)\hat{\sigma}^2+\frac{n}{2}h_n^2(\hat{\sigma}^2_B+\hat{\sigma}^2_A)]=\]
\[
\frac{1}{n}[(1 +h_n)\hat{\sigma}^2+\frac{1}{2}h_n^2(\hat{\sigma}^2_B+\hat{\sigma}^2_A)]\]
Thus the asymptotic variance is
\[
VAR[\sqrt{n}\hat{\Tilde{d}}] = nVAR[\hat{\Tilde{d}}]=(1 +h_n)\hat{\sigma}^2+\frac{h_n^2}{2}(\hat{\sigma}^2_A+\hat{\sigma}^2_B)=\hat{\Tilde{\sigma}}^2\]

\section{Vectorized form of ODE derivatives within SW-test framework}
\label{vecderiv}
Our testing approach applied to systems of ODE is novel. Moreover, there is no ready to use software that produces automatically required Jacobian and Hessian. The Python implementation we developed is based on the "casadi" package that provides the first and the second numerical derivatives of a system of ODE with respect to $\xi$ and $\psi$ as separate matrices. We need to use these blocks to construct the final full Hessian and Jacobian with respect to $\theta$ combining them with parts obtained from taking derivatives over the log of density. Our goal is to derive the Jacobian:
\[\nabla_\theta ln(f(y|t, \theta)) = \begin{bmatrix}
\frac{\partial ln(f(y|t_i, \theta))}{\partial \sigma^2} &
\frac{\partial ln(f(y|t_i, \theta))}{\partial \xi}
&
\frac{\partial ln(f(y|t_i, \theta))}{\partial \psi}
\end{bmatrix}    
\]
and Hessian:
\[
\nabla_\theta^2 ln(f(y|t_i, \theta)) = 
\begin{bmatrix}
\frac{\partial^2 ln(f(y|t_i, \theta))}{\partial (\sigma^2)^2} &
\frac{\partial^2 ln(f(y|t_i, \theta))}{\partial \sigma^2 \partial \xi}& 
\frac{\partial^2 ln(f(y|t_i, \theta))}{\partial \sigma^2 \partial \psi}
\\
\frac{\partial^2 ln(f(y|t_i, \theta))}{\partial \xi \partial \sigma^2}&
\frac{\partial^2 ln(f(y|t_i, \theta))}{\partial (\xi)^2}& 
\frac{\partial^2 ln(f(y|t_i, \theta))}{\partial \xi \partial \psi}
\\
\frac{\partial^2 ln(f(y|t_i, \theta))}{\partial \psi \partial \sigma^2}& 
\frac{\partial^2 ln(f(y|t_i, \theta))}{\partial \psi \partial \xi}& 
\frac{\partial^2 ln(f(y|t_i, \theta))}{\partial (\psi)^2}
\end{bmatrix}
\]
where $ln(f(y|t, \theta))$ is the logarithm of the joint density of the assumed distribution. Following the main text we consider Gaussian noise with zero means and finite $\sigma^2_j, j=1\dots d$, and thus the log-density of a single observation is:
\[
\ln (f(y_i|t, \theta)) = -\frac{1}{2}\log(2\pi) - \frac{1}{2}\log(\prod_j^d\sigma_j^2)-\sum_j^d\frac{(y_{i,j}-x_j(t_i, \eta))^2}{2\sigma^2_j}.
\]

To save space we will use in the sequel $t$ instead of $t_i$ to define an arbitrary data point. In the following subsections, we provide quite a detailed way of obtaining derivatives. To make the vectorized form of calculating derivatives provided in Section \ref{der_general} more clear we will start with a specific example showing how derivatives can be calculated directly (knowing the exact solution of ODE system) and by vectorization using the Jacobian and Hessian defined above.

\subsection{Example of calculating the derivatives}
\label{derivexample}
Consider the following system of exponential decay equations with three states and two parameters:
\[
\left\{\begin{matrix}
 x_1^\prime(t)=&\psi_1 x_1(t)\\
 x_2^\prime(t)=&\psi_2 x_2(t)\\
 x_3^\prime(t)=&(\psi_1 + \psi_2) x_3(t)\\
\end{matrix}\right.
\]
The solution to the system is the following:
\[
\left\{\begin{matrix}
x_1(t,\eta)=&\xi_1\mathrm{e}^{\psi_1 t} \\
x_2(t,\eta)=&\xi_2\mathrm{e}^{\psi_2 t}\\
x_3(t,\eta)=&\xi_3\mathrm{e}^{(\psi_1+\psi_2) t}
\end{matrix}\right.
\]
Consider the regression model studied in the main text, namely that we observe the solution of differential equation with an additive Gaussian measurement error. Then the log-density of a single observation is given by:
\[
\ln (f(y_i|t, \theta)) = -\frac{1}{2}\log(2\pi) - \frac{1}{2}\log(\sigma_1^2\sigma_2^2\sigma_3^2)-\frac{(y_{i,1}-x_1(t, \eta))^2}{2\sigma^2_1}-\frac{(y_{i,2}-x_2(t, \eta))^2}{2\sigma^2_2}-\frac{(y_{i,3}-x_3(t, \eta))^2}{2\sigma^2_3}.
\]
According to Equation \ref{7.1} below the general form of the Jacobian for this example is the following
\[
\nabla_\theta ln(f(y|t, \theta)) = 
.
\]
Using the solution of the ODE system and the log density we can directly derive that:
\begin{equation}
\label{der_J_exam}
\nabla_\theta ln(f(y|t, \theta)) = %
^{\top}
\end{equation}

According to Equation \ref{der_H} the Hessian for this example is the following:

\[
\nabla_\theta^2 ln(f(y|t, \theta)) = 
%
.
\]
In our specific example, the first row of the Hessian matrix is given by:
\small
\begin{equation}
\label{der_H_log_sigma_exam}
%
^{\top}
\end{equation}
The second row of the Hessian matrix takes the form

\begin{equation}
\label{der_H_log_xi_exam}
%
^{\top}
\end{equation}
And the third row of the above Hessian matrix is

\tiny
\begin{equation}
\label{der_H_log_psi_exam}
%
^{\top}
\end{equation}
\normalsize
\\

In most cases, the exact solution for the ODE system is impossible to derive, so one can use the vectorized form described in Section \ref{vecderiv}. Let's apply it to the current example and check if its output coincides with the above Jacobian and Hessian calculated directly.\\

To move forward, let's derive the required vectors and matrices:
\[
\mathbf{g}^{(i)}=%
(see \ Equation \ \ref{der_D_psi})
\]

and according to Equation \ref{der_J}:
\[
\nabla_\theta ln(f(y|t, \theta)) = %
\]
Let's calculate each element separately:
\[
\frac{1}{2}((\mathbf{g}^{(i)})^2*\mathbf{\frac{1}{\sigma^2}}-1)*\mathbf{\frac{1}{\sigma^2}}=
\frac{1}{2}(%
\]
that is equivalent to the first three rows of the vector \ref{der_J_exam} (note that it is transposed).\\

Next:
\[
[\mathbf{D}^{(i)}_{\xi}(\mathbf{\frac{1}{\sigma^2}}*\mathbf{g}^{(i)})^{\top}]^{\top}=
[%
\]
that is equivalent to the 4,5,6th rows of the vector \ref{der_J_exam}.\\

Next:
\[
[\mathbf{D}^{(i)}_{\psi}(\mathbf{\frac{1}{\sigma^2}}*\mathbf{g}^{(i)})^{\top}]^{\top}=
[%
\]
that is equivalent to the last rows of the vector \ref{der_J_exam}.\\

To move forward with Hessian we need to derive auxiliary vectors:
\[
%
\]
 (see \ Equation \ \ref{der_H_v_xi2} for the above).
\[
%
\]
(see \ Equation \ \ref{der_H_v_xi_xi} for the above)
 
\[
%
\]
(see \ Equation \ \ref{der_H_v_xi} for the above).
\[
%
\]
(see \ Equation \ \ref{der_H_v_xi_psi} for the above).
\[
%
\]
(see \ Equation \ \ref{der_H_v_psi} for the above).
\[
%
\]
(see \ Equation \ \ref{der_H_v_psi_xi} for the above).
\[
%
\]
(see \ Equation \ \ref{der_H_v_psi_psi} for the above).
\[
\mathbf{v_{\frac{g^{(i)}}{\sigma^2}}}=
%
\]
Now we can derive each element of the Hessian matrix (see Equation \ref{der_H_vec}). According to Equation \ref{der_H_sigma}:
\[
(\frac{\partial ln(f(y|t, \theta))}{\partial\sigma^2})^{'}_{\theta} = 
%
^{\top}
\]
where for this example:
\[
\Delta(\frac{1}{2}(\mathbf{\frac{1}{\sigma^2}})^2 - (\mathbf{g}^{(i)})^2*(\mathbf{\frac{1}{\sigma^2}})^3) = 
%
\]
that coincides with the first 3 rows of Equation \ref{der_H_log_sigma_exam};
\[
 -[\mathbf{D}^{(i)}_{\xi}\Delta(\mathbf{g}^{(i)}*(\mathbf{\frac{1}{\sigma^2}})^2)]^{\top} = 
-[%
\]
that coincides with the 4, 5, 6th rows of Equation \ref{der_H_log_sigma_exam};
\[
 -[\mathbf{D}^{(i)}_{\psi}\Delta(\mathbf{g}^{(i)}*(\mathbf{\frac{1}{\sigma^2}})^2)]^{\top} = 
-[%
\]
that coincides with the last two rows of Equation \ref{der_H_log_sigma_exam}. Thus we can conclude that in this example the result of using vectorization (Equation \ref{der_H_sigma}) is equivalent to the result obtained via exact solution to the given system of ODEs (Equation \ref{der_H_log_sigma_exam}).

Next, according to Equation \ref{der_H_xi}:
\[
%
    \]
    that coincides with the first three rows of Equation \ref{der_H_log_xi_exam};    
    
    \item  (see Equation \ref{der_log_H_xi2})
    \[
    \frac{\partial^2 ln(f(y|t, \theta))}{\partial (\xi)^2} =
    \mathbf{V}_{2}^{(\xi)}-
    \mathbf{V}_{1}^{(\xi)} = 
    %
    \]
    that coincides with 4,5,6th rows of Equation \ref{der_H_log_xi_exam}. For details see Equations \ref{der_V_xi_2}, \ref{der_V_xi_1}:
    \[
    \mathbf{V}_{2}^{(\xi)} = %
    \]
    that coincides with the last two rows of Equation \ref{der_H_log_xi_exam}. For details see Equations \ref{der_V_xi_psi_2}, \ref{der_V_xi_psi_1}:
    \[
    \mathbf{V}_{2}^{(\xi,\psi)} = %
    \]
\end{itemize}
The following can be derived using the same concepts as for $(\frac{\nabla_\theta ln(f(y|t, \theta))}{\nabla \xi})^{'}_{\theta}$:
\[
%
=
    \]
    that coincides with the first three rows of Equation \ref{der_H_log_psi_exam};    
    \item 
     \[
    \frac{\partial^2 ln(f(y|t, \theta))}{\partial \psi \partial \xi} =
    \mathbf{V}_{2}^{(\psi,\xi)}-
    \mathbf{V}_{1}^{(\psi,\xi)} = 
    \]
    \[
    %
    \]
    that coincides with 4,5,6th rows of Equation \ref{der_H_log_psi_exam}, and where:
    \[
    \mathbf{V}_{2}^{(\xi,\psi)} = %
    \]
    \normalsize
    that coincides with the last two rows of Equation \ref{der_H_log_psi_exam} and where:
    \[
    \mathbf{V}_{2}^{(\psi)} = %
    \]
    
\end{itemize}

As we see, with separate blocks of derivatives we can obtain the same Jacobian and Hessian as if we knew the exact solution of ODE system.

\subsubsection{Generalization of calculating the derivatives}
\label{der_general}
The compact form of $\nabla_\theta ln(f(y|t, \theta))$ can be represented as follows:
\begin{equation}
\label{7.1}
\nabla_\theta ln(f(y|t, \theta)) = %
,    
\end{equation}

where $\theta$ is defined according to 
\begin{equation}\nonmuber
\theta=(\sigma^2, \xi, \psi),
\end{equation}
and $\frac{\partial ln(f(y|t, \theta))}{\partial \sigma^2}$ and $\frac{\partial ln(f(y|t, \theta))}{\partial \xi}$ are vectors of size $1\times d$ and $\frac{\partial ln(f(y|t, \theta))}{\partial \psi}$ is a vector of size $1\times p$, stacked into $1\times (2d+p)$ row-vector.\\

In the Gaussian additive error case, the specific terms of $\nabla_\theta ln(f(y|t, \theta))$ are as follows:
\[
\nabla_\theta ln(f(y|t, \theta)) = %
^{\top}
\]

By rearranging the terms, we get

a)
\[
\frac{\partial ln(f(y|t, \theta))}{\partial \sigma^2} =\frac{1}{2}((\mathbf{g}^{(i)})^2*\mathbf{\frac{1}{\sigma^2}}-1)*\mathbf{\frac{1}{\sigma^2}}
\]
Here $*$ defines elementwise multiplication of two vectors/matrices ($c_{i,j}=a_{i,j}*b_{i,j}$), $\mathbf{g}^{(i)}$ is a vector of $(y_{i,j}-x_j(t, \eta))$ of dimension $1\times d$:
\begin{equation}
\label{der_g}
\mathbf{g}^{(i)}=%
    
\end{equation}

b)
\[
\frac{\partial ln(f(y|t, \theta))}{\partial \xi}=[\mathbf{D}^{(i)}_{\xi}(\mathbf{\frac{1}{\sigma^2}}*\mathbf{g}^{(i)})^{\top}]^{\top}
\]
Here $\mathbf{D}^{(i)}_{\xi}$ is a matrix of size $d\times d$ of partial derivatives of $x(t, \eta)$ with respect to $\xi$:
\begin{equation}
\label{der_D_xi}
\mathbf{D}^{(i)}_{\xi}=
%
\end{equation}

c)
\[
\frac{\partial ln(f(y|t, \theta))}{\partial \psi}=[\mathbf{D}^{(i)}_{\psi}(\mathbf{\frac{1}{\sigma^2}}*\mathbf{g}^{(i)})^{\top}]^{\top}
\]
where $\mathbf{D}^{(i)}_{\psi}$ is a matrix of size $p\times d$ of partial derivative of $x(t, \eta)$ with respect to $\psi$:

\begin{equation}
\label{der_D_psi}
\mathbf{D}^{(i)}_{\psi}=
%
\end{equation}

As a result, we have
\begin{equation}
\label{der_J}
\nabla_\theta ln(f(y|t, \theta)) = %
    
\end{equation}

Next, we derive the general form of the Hessian:
\[
\nabla_\theta^2 ln(f(y|t, \theta)) = 
%
\end{equation}
Consider each part at a time:
\[
(\frac{\partial ln(f(y|t, \theta))}{\partial \sigma^2})^{'}_{\theta} = 
\]
\[
%
\end{equation}
where $\Delta()$ is the function that converts an arbitrary vector $\mathbf{v}$ into the diagonal matrix M, such that $M_{i,i}=\mathbf{v}_i, M_{i,j\neq i}=0$. For example:
\begin{equation}
\label{der_delta_sigma}
\Delta(\mathbf{g}^{(i)}*(\mathbf{\frac{1}{\sigma^2}}))=
%
\end{equation}

Next, we consider:
\begin{equation}
\label{der_H_xi}
%
=
\]
Let's $\mathbf{v}_{\xi_s\xi_r}$ be a column vector of the second derivatives of the vector of functions $x(t, \eta)$ with respect to a pair $\xi_s\xi_r, (s,r\in(1,2,...,d))$. For example:
\begin{equation}
\label{der_H_v_xi2}
\mathbf{v}_{(\xi_1)^2}=
%
^{\top}
\end{equation}
And let $\mathbf{v}_{\partial \xi_s}$ be a column vector of the first derivatives of the vector of functions $x(t, \eta)$ with respect to $\xi_s (s\in(1,2,...,d))$. For example:
\begin{equation}
\label{der_H_v_xi}
\mathbf{v}_{\xi_1} = 
%
^{\top}
\end{equation}
Now we can define the following:

\begin{equation}
\label{der_V_xi_2}
\mathbf{V}_{2}^{(\xi)} = %
\end{equation}
where vector $\mathbf{v_{\frac{g^{(i)}}{\sigma^2}}}=\mathbf{g}^{(i)}*\mathbf{\frac{1}{\sigma^2}}$, and vector $\mathbf{v_{\frac{1}{\sigma^2}}}=\mathbf{\frac{1}{\sigma^2}}$. Combining them all together we derive that:
\begin{equation}
\label{der_log_H_xi2}
\frac{\partial^2 ln(f(y|t, \theta))}{\partial (\xi)^2} = \mathbf{V}_{2}^{(\xi)}-
\mathbf{V}_{1}^{(\xi)}
\end{equation}

c)
\[
\frac{\partial^2 ln(f(y|t, \theta))}{\partial \xi \partial \psi} =
\]
\[
%
\end{equation}
where $\mathbf{v}_{\xi_s \psi_q}$ is a column vector of the second derivatives of the vector of functions $x(t, \eta)$ with respect to a pair $\xi_s\psi_q, (s\in(1,2,...,d), q\in(1,2,...,p))$:
\begin{equation}
\label{der_H_v_xi_psi}
\mathbf{v}_{\xi_s \psi_q}=
%
^{\top}
\end{equation}
And $\mathbf{v}_{\psi_q}$ is a column vector of the first derivatives of the vector of functions $x(t, \eta)$ with respect to $\psi_q (q\in(1,2,...,p))$:
\begin{equation}
\label{der_H_v_psi}
\mathbf{v}_{\psi_q} = 
%
\]

By the same token as for $(\frac{\partial ln(f(y|t, \theta))}{\partial \xi})^{'}_{\theta}$, we can derive that:
\[
(\frac{\partial ln(f(y|t, \theta))}{\partial \psi})^{'}_{\theta} = 
%
\]
where $\mathbf{V}_{2}^{(\psi)}, \mathbf{V}_{1}^{(\psi)}, \mathbf{V}_{2}^{(\psi, \xi)}, \mathbf{V}_{1}^{(\psi, \xi)}$ are constructed in the same way as $\mathbf{V}_{2}^{(\xi)}, \mathbf{V}_{1}^{(\xi)}, \mathbf{V}_{2}^{(\xi, \psi)}, \mathbf{V}_{1}^{(\xi, \psi)}$, but with respect to $\psi$ at first:\\
\begin{equation}
\label{der_H_v_psi_xi}
\mathbf{v}_{\psi_q \xi_s}=
%
^{\top}
\end{equation}
where $q, m\in(1,2,...,p), s\in(1,2,...,d)$.

Finally, we conclude that:
\begin{equation}
\label{der_H_vec}
%
\end{equation}\\

The final and important note is that one should pay attention to the form of handling the order of parameters in the output of the software (by rows (left) or by columns (right)):
\[
%
\]
Our representation is based on the "row-order", so if the user's software provides "column-order" output all equations should be transposed correspondingly.

\bibliography{main}